\documentclass[twocolumn,aps]{revtex4}
\usepackage{graphicx}

\begin{document}
\title{A Coherent View on Entropy}

\author{Won-Young Hwang \footnote{wyhwang@jnu.ac.kr}
}

\affiliation{ Department of Physics Education, Chonnam National
University, Gwangju 500-757, Republic of Korea
}

\begin{abstract}
Informational entropy is often identified as physical entropy. This is surprising because the two quantities are differently defined and furthermore the former is a subjective quantity while the latter is an objective one. We describe the problems and then present a possible view that reconciles the two entropies. Informational entropy of a system is interpreted as physical entropy of a whole composed of both the system and ``memories" containing information about the system.
\vspace{2mm}

\noindent{PACS:}
05.30.-d, 05.20.-y, 03.67.-a \\
Keywords: Physical Entropy, Information Entropy, Memory of Observer
\end{abstract}

\maketitle
\section{Introduction}\label{sec:intro}
The second law of thermodynamics is one of the most concrete laws in nature \cite{Wal85,Rei65,Val91,Per95}. A quantity $S$, physical entropy (or thermodynamic entropy), of an isolated system does not decrease. Boltzmann gave a microscopic interpretation of the physical entropy \cite{Wal85,Rei65},
\begin{equation}
\label{1}
 S=\log W.
\end{equation}
Here $W$ is the number of microstates belonging to a given macrostate. (We set  $k \hspace{0.5mm} ln=log_2=log$ where $k$ is Boltzmann constant.) Let us consider physical entropy of a microstate. (It is often said that physical entropy cannot be assigned to a microstate. But this is a misconception. A physical system is in a  ``microstate" not in a ``macrostate" at any time.) At each instant, the physical system must be in a ``microstate" corresponding to a macrostate of the physical system.  Physical entropy of the microstate should be that of the macrostate which the microstate belongs to. That is, the physical entropy $S$ of a microstate is defined by the number of microstates belonging to the same macrostate. (The next task is to appropriately divide the phase space such that each region of the phase space corresponds to a macrostate.  This is a very subtle problem which is not topic of this paper.) Clearly, the physical entropy $S$ is an objective quantity, and is observer independent.  A physical system in a microstate belonging to a macrostate has a fixed value of $S$, whether one knows identity of the microstate or not.

On the other hand, Shannon proposed a quantity $H$, informational entropy,
\begin{equation}
\label{2}
 H=-\sum_i p_i \log p_i.
\end{equation}
to measure randomness of an entity \cite{Sha48}. Here $p_i$ denotes the probability that the entity will be the $i$-th. It should be emphasized that informational entropy $H$ is a subjectively defined quantity.  When we talk about probability, we tacitly assume ``someone" for whom the probability is defined. In order that probability be meaningful, we should clarify whose probability it is.

{\bf Proposition 1}: {\it Informational entropy $H$ is meaningless without an observer for whom the probability $p_i$'s are defined.}

Here we denoted  the ``someone" by an ``observer".

Surprisingly, however, informational entropy $H$ is often identified with physical entropy $S$ \cite{Wal85,Rei65,Val91,Per95,Sha48,Sch10}. On the one hand, the two quantities are not expected to be the same because the two quantities are differently defined and furthermore the former is a subjective quantity while the latter is an objective one. On the other hand, however, there can be a deeper reason to justify the identification \cite{Sch10}.  The purpose of this paper is to examine the identification and provide a coherent explanation; Informational entropy $H$ of a system can be interpreted as physical entropy $S$ of the whole including both the system and ``memories of outside observer" correlated with the system.

This paper is organized as follows. Section II gives an explanation about why the two quantities are not to be simply identified.  In section III, we discuss explanations proposed so far.  In section IV, the main part of this paper, we describe how informational entropy can be interpreted in terms of objective entropy. In section V, a related question is discussed. The conclusion is reported in section VI.
\section{Informational Entropy and Physical Entropy Are Differently Defined: Entropy of Universe?}
As introduced previously, informational entropy and physical entropy are subjective and objective quantity, respectively, being differently defined. Thus the two quantities are not expected to be the same.  Here we consider a question, ``entropy of universe?", which illustrates difficulty of the identification.

The same problem exists in both classical and quantum mechanics. Here we pose the problem in a quantum form that might be more familiar to physicists.  Von Neumann introduced a quantum analog of informational entropy,
\begin{equation}
\label{3}
N= -\sum_i p_i \log p_i,
\end{equation}
for a density operator $\rho= \sum_i p_i |i\rangle \langle i|$ where $|i\rangle$'s are orthonormal states and  $p_i$'s are associated probabilities. The density operator corresponds to an ensemble of the quantum states $|i\rangle$ with a probability  $p_i$. Whether a state is pure or mixed is observer dependent. A state that is pure to one observer can be mixed to another. However, von Neumann entropy of a pure state and mixed one are different. The former is zero while the latter is positive. Therefore, von Neumann entropy is observer dependent. This is as expected because von Neumann entropy is a type of informational entropy.

It is difficult to expect that an observer dependent quantity, von Neumann entropy, can be simply identified with an objective quantity, physical entropy.  Let us consider a question ``What is the (von Neumann) entropy of the universe?". One might say that entropy of universe is zero because the universe is in a (huge) pure state. This is in contradiction with a fact that physical entropy of universe is enormous.  A correct answer to the question is that von Neumann entropy is not defined for the universe. Recall that there is no one outside the universe for whom the probability $p_i$ is defined, and Proposition 1.

However, it should be that physical entropy can be defined for the universe.  If a quantity can be defined but the other one cannot be for an identical case, clearly we cannot simply identify the two quantities.
\section{Explanations Proposed So Far}
A common explanation about the identification is as follows. Consider a macrostate composed of $W$ different microstates.  Assume that an observer has no knowledge about microstate. In this case we should assume equal probability for each microstate. Then we get   $p_i=1/W$, with which informational entropy in Eq. (\ref{2}) reduces to physical entropy in Eq. (\ref{1}).

However, the argument implies only that the identification is valid in the specific case when the observer has zero knowledge. The argument does not justify usual identification of information entropy with physical entropy in a general sense. Suppose the observer somehow knows in which microstate it is. The informational entropy becomes zero because all $p_i$ except for one are zero. But the physical entropy is unchanged even in this case.

In another interesting argument for the identification, it is utilized that mixing process increases physical entropy \cite{Per95}.  Suppose that two chambers of identical volume $V$ are filled with gas $A$ and $B$, respectively. (For simplicity, here we consider the special classical case although the argument is valid for more general and quantum cases.) Then we mix gases by connecting the chambers. Using semi-permeable membrane, it can be shown that physical entropy increase per molecule involved with the mixing process is $\log 2$ \cite{Per95, Mar09}. Now let us check informational entropy change. Before mixing, observer knows whether a specified molecule is $A$ or $B$, once the observer knows in which chamber the molecule is. After mixing, observer knows nothing about identity of a specified molecule because all molecules are mixed in a chamber of volume $2V$. Thus informational entropy per molecule before and after mixing is 0 and $\log 2$, respectively. Therefore, change of informational entropy due to the mixing is $\log 2$, which coincides with the change of physical entropy.

What we can see by the argument is, however, that the two quantities can be identified for the mixing process. The argument doesn't say something about other cases. For example, let us consider a hypothetical case in which observer knows position and velocity of each molecule although all molecules are mixed in the chamber of volume $2V$. In this case, informational entropy is $0$ because observe has full knowledge of each molecule. The result is confusing because the gases are fully mixed. The argument in Ref. \cite{Per95} does not provide an explanation for this case.
\section{A coherent explanation:  H (system)= S(system +memory)}
The paradox is the followings. Suppose observer somehow knows in which microstate a macrostate is. Then informational entropy is zero because all $p_i$ except for one are zero. Does this mean physical entropy is also zero? Physical entropy should be non-zero by Eq. (\ref{1}). However, there is an intuitive reasoning claiming that it should be zero. Because the observer has full information about a physical system (the macrostate), the observer can manipulate the system as desired. For example, suppose an observer knows positions and velocities of all gas molecules in a chamber which appear to be randomly distributed. For the observer with the knowledge, however, the gas is no longer randomly distributed. In principle the observer can align motions of all molecules, as a result heat of the gas is converted to a lump of gas's kinetic energy. Therefore, knowledge indeed decreases physical entropy.

A key fact in the argument is that there is ``someone" or an ``observer" who will use the knowledge. This suggests that the Proposition 1 can be a clue for resolution of the paradox.  Here let us consider  ``memories" of the observer. We can see that the more knowledge about the molecules the observer has, the more correlated with the molecules the memories are.  We consider not only a system (the molecules) but also the observer's memories.  However, ``more correlation" between molecules and memories means ``less volume" in phase space occupied by the combined whole of molecules and memories. ``Less volume" then implies ``less physical entropy" by Eq. (\ref{1}).

{\bf Explanation 1}: {\it Informational entropy of a system can be interpreted as physical entropy of both the system and memories containing the information on the system, namely $H$ (system)=$S$(system+memory)}.

Let us illustrate Explanation 1. We consider informational entropy $H$ of a system of (a positive integer)  $n$  quantum bits (qubits). (It doesn't matter if bits are classical or quantum. The same thing can be said for classical bits. Also we omit explanation for intermediate cases for simplicity.)  For each qubit, there is a corresponding qubit, which plays the role of memory. There will be $n$  memory qubits. Here we consider only those cases where each qubit is either in one of two orthogonal states, $|0\rangle$ and $|1\rangle$. Two extreme cases are considered.

The first case is when the system and memory are perfectly correlated. This means that memory has perfect information about the system. In this case, each qubit of the system might be either $|0\rangle$ or $|1\rangle$. However, each memory-qubit is in the same state as that of the corresponding system-qubit. (One may assume that the system-qubit and memory-qubit is anti-correlated, which makes no difference to our arguments.) For example, $(|0\rangle,|0\rangle),(|1\rangle,|1\rangle),(|0\rangle,|0\rangle),
(|0\rangle,|0\rangle),(|1\rangle,|1\rangle)$,... . Here the first (second) one in each parenthesis denotes the system-qubit (memory-qubit). Therefore, the total number of states $W$ is  $2^n$. From Eq. (\ref{1}), we have
\begin{equation}
\label{4}
S=\log 2^n= n.
\end{equation}
Informational entropy in this case is zero because memory gives full information about the system.
\begin{equation}
\label{5}
H= 0.
\end{equation}
The second case is when the system and memory are not correlated. This means that memory has no information about the system. For example, $(|0\rangle,|1\rangle),(|1\rangle,|1\rangle),(|0\rangle,|1\rangle),
(|0\rangle,|0\rangle),(|1\rangle,|0\rangle)$,... . The total number of states $W$ is $2^{2n}$ and thus
\begin{equation}
\label{6}
 S^{\prime}=\log 2^{2n}= 2n.
\end{equation}
Informational entropy in the second case is easily derived from $p_i=1/2^n$  and Eq. (\ref{2}),
\begin{equation}
\label{7}
 H^{\prime}= n.
\end{equation}
Therefore, difference of physical entropy is equal to that of informational entropy.
\begin{equation}
\label{8}
S^{\prime} -S= H^{\prime} -H= n.
\end{equation}
Informational entropy change due to loss of information can now be explained in terms of physical entropy change of the combined whole of system and memory.
\section{A related problem}
A related problem involved with the identification, often regarded as a paradox, is ``the (informational) entropy is constant \cite{Val91,Weh78}". (Similar problem exists in classical mechanics. Due to the Liouville theorem, the (informational) entropy is constant.) Because the Schrodinger equation is deterministic, the observer can track the quantum state. If one knows the initial quantum state then the one knows quantum state later. The fact that it is possible to track and unitarity of quantum mechanics mean that the informational entropy $N$ is constant. This is the origin of the famous problem involved with blackhole evaporation \cite{Zeh05,Haw04}.  However, as discussed previously, if the system is universe, informational entropy is meaningless because there is no observer outside universe.  So we should assume that the system is a part of the universe. Here we identify informational entropy with physical entropy of a whole composed of the system and another part corresponding to memories. In order to show that the informational entropy of the system increases, we need to show that physical entropy of the whole increases. The problem needs to be studied later. However, intuitively speaking, correlation between the system and memory will degrades (decreases) in time and thus the entropy increases. This explanation might be related to resolution of the paradox by the ``coarse graining" idea \cite{Val91,Wal85}.
\section{conclusion}
One of the most puzzling questions regarding entropy is that informational entropy is often identified as physical entropy. It is because the two quantities are differently defined. The former is a subjective quantity while the latter is an objective one. There are a few previous explanations (or interpretations) for the paradox but these are meaningful in some specified contexts. Here we provide an interesting explanation which, we find, reveals the core of the problem. We made a key observation that informational entropy is not defined without ``observer". (This implies that informational entropy of the universe is not defined.) Based on the observation, we presented a possible view that reconciles the two entropies. Informational entropy of a system is interpreted as physical entropy of a whole composed of both the system and ``memories (of observer)" containing information about the system. We expect that our explanation will shed light on resolution of the paradox.
\section*{Acknowledgement}
This research was supported by Basic Science Research Program through the National Research Foundation of Korea (NRF) funded by the Minstry of Education, Science and Technology (2010-0007208).

\end{document}